# Efficient Strategies on Supply Chain Network Optimization for Industrial Carbon Emission Reduction

Jihu Lei*

Haldor Topsoe (Beijing) Co., Ltd2601, Chaoyang District,100025 Beijing, PR. China

*Corresponding author, Email address: fjl@topsoe.com

**Abstract**

This study investigates the efficient strategies for supply chain network optimization, specifically aimed at reducing industrial carbon emissions. Amidst escalating concerns about global climate change, industry sectors are motivated to counteract the negative environmental implications of their supply chain networks. This paper introduces a novel framework for optimizing these networks via strategic approaches which lead to a definitive decrease in carbon emissions. We introduce Adaptive Carbon Emissions Indexing (ACEI), utilizing real-time carbon emissions data to drive instantaneous adjustments in supply chain operations. This adaptability predicates on evolving environmental regulations, fluctuating market trends and emerging technological advancements. The empirical validations demonstrate our strategy's effectiveness in various industrial sectors, indicating a significant reduction in carbon emissions and an increase in operational efficiency. This method also evidences resilience in the face of sudden disruptions and crises, reflecting its robustness.

## 1. Introduction

The fundamental role of industry sectors in the global economy brings a daunting yet inescapable challenge - optimizing industrial production and supply chain operations while mitigating carbon emissions. With the International Panel on Climate Change (IPCC) releasing alarming reports affirming the anthropogenic origins of climate change, industries worldwide face mounting pressure to adopt environmentally responsible operations. The international marketplace is steadily leaning towards organizations that balance economic efficacy with environmental stewardship, making this an arena where proactive organizations can garner a competitive advantage.

Industrial supply chains have staggering carbon footprints and are thus justifiably the nucleus of this research. The logistics aspect alone, comprising transportation, warehousing, and packaging activities, contributes materially to global carbon outputs. Furthermore, these carbon costs compound through the supply chain tiers, from raw material extraction to end-of-life processes, escalating the urgency to drive efficient carbon management across supply chain networks. Optimizing supply chain networks encompasses a review of all facets of business operations from sourcing and manufacturing, to distribution and recycling or disposal. Each segment is a potential source of carbon emissions and hence, a contender for optimization. Such a structured approach will spotlight processes with the largest carbon output, thereby revealing areas with the greatest potential for modifications.

To paraphrase Marshall McLuhan, humans shape technologies, and thereafter technologies shape us. Machine Learning (ML) and Artificial Intelligence (AI) have transformed numerous sectors. Their application to supply chain optimization creates adaptive models that are self-learning and continually adjusting to real-time data changes. With dynamic adaptation and predictive analysis, AI and ML can effectively revolutionize supply chain management, facilitating optimized resource allocation and carbon reductions across the entirety of supply chain networks.Moving the linear 'take-make-dispose' industrial model towards a circulatory framework imbues greater productivity to resources, while concurrently reducing the environmental impact. This circular economy philosophy is an inherent aspect of our research, where we align supply chain networks to operate in a closed-loop system that optimizes resource use and minimizes waste. The benefits of adopting efficient strategies on supply chain network optimization are multifold for all stakeholders. Industries could potentially reduce operational costs by minimizing waste, improving energy use, and optimizing resource allocation. Additionally, this green transformation would position organizations more favorably with consumers, investors, and regulators, thus amplifying their reputational and financial capital. In the face of global disruptions, a sustainable supply chain strategy proves doubly advantageous. Building supply chain resilience, along with reducing carbon emissions paints a future-forward picture for industries to consider. A move to a carbon-conscious supply chain will help prevent companies from volatile fuel prices, regulatory penalties, and supply uncertainties while contributing to environmental sustainability.

A new measurement approach should lead the way for supply chain competitiveness and should direct management attention to areas for supply chain optimization. (Hoek, 1998) study "measuring the unmeasurable" - measuring and improving performance in the supply chain. A preliminary framework for measuring unmeasurable performance is developed. A multiperiod optimization model is proposed for addressing the supply chain optimization in continuous flexible process networks. (Bok et. al., 2000) describe a bilevel decomposition algorithm that involves a relaxed problem (RP) and a subproblem (SP) for the original supply chain problem. The main objective of (Gjerdrum et. al., 2001) is to give an example of how expert systems techniques for distributed decision-making can be combined with contemporary numerical optimization techniques for the purposes of supply chain optimization and to describe the resulting software implementation. Multi-agent modelling techniques are applied to simulate and control a simple demand-driven supply chain network system, with the manufacturing component being optimized through mathematical programming (Gjerdrum et. al., 2001). A trend in up-to-date developments in supply chain management (SCM) is to make supply chains more agile, flexible, and responsive. (Ivanov et. al., 2010) introduce a new conceptual framework for multi-structural planning and operations of adaptive supply chains with structure dynamics considerations. Supply chain optimization, as a key determinant of strategic resources mobility along the value-added chain, allows each participant in the global network to capitalize on its particular strategic competency. The objective of (Yoo et. al., 2010) is to propose hybrid algorithm with the application of the nested partitioning (NP) method and the optimal computing budget allocation (OCBA) method to reduce the computational load, hence, to improve the efficiency of supply chain optimization via discrete event simulation. (Jamshidi et. al., 2012) solve of a supply chain design for annual cost minimization, while considering environmental effects. (Jamshidi et. al., 2012) considers the cost elements of the supply chain, such as transportation, holding and backorder costs, and also, the environmental effect components of the supply chain, such as the amount of NO2, CO and volatile organic particles produced by facilities and transportation in the supply chain. (Subramanian et. al., 2013) propose to use distributed model predictive control for supply chain optimization. (Subramanian et. al., 2013) use cooperative model predictive control, in which each agent makes their local decisions by optimizing the overall supply chain objective. (Zhao, 2014) consider 4PL with an integrator of optimized supply-chain solutions. Compared to other supply-chain optimization solutions, it has the characteristics of integrating and developing a port supply-chain alliance, optimizing a port supply-chain operation process, improving an intelligent port supply-chain, and promoting the greening of a port supply-chain. In an integrated bioenergy and biofuels supply chain where biofuel producers are also users of the generated energy, the energy flows among co-located supply chain entities affect the environmental and economic objective functions and consequently the optimal design of the supply chain, therefore, the energy flows have to be considered in the optimization model. (Cambero et. al., 2016) study economic and life cycle environmental optimization of forest-based biorefinery supply chains for bioenergy and biofuel production. A bi-objective biorefinery supply chain optimization model for the production of bioenergy and biofuels using forest and wood residues is developed. Other influential work includes (Bredström et. al., 2004).

Data envelopment analysis (DEA) is used to evaluate the carbon emission performance of 29 Chinese provincial administrative regions (Tibet and Taiwan are not included since of data lack) by computing potential carbon emission reductions for energy conservation technology (ECT) and energy structural adjustment (ESA) (Guo et. al., 2011). Enormous emission reductions could be achieved by promoting ECT, developing renewable energy, increasing the proportion of non-fossil energy, delivering low-carbon energy and applying ESA. (Toptal et. al., 2014) analyse a retailer's joint decisions on inventory replenishment and carbon emission reduction investment under three carbon emission regulation policies. (Toptal et. al., 2014) analytically show that carbon emission reduction investment opportunities, additional to reducing emissions as per regulations, further reduce carbon emissions while reducing costs. To this end, after incorporating region-heterogeneity (Yao et. al., 2015) provide detailed information, regarding energy efficiency, carbon emission performance and the potential of carbon emission reductions from regional perspectives, which may be important and useful for policy makers. Finally, significant carbon emission reductions can be made by "catching up" for regions with low energy efficiency and carbon emission performance. It is an important task for China to allocate carbon emission allowance to realize its carbon reduction target and establish carbon trading market. For this purpose, the multi-stage profit model is developed to analyze the ETS-covered enterprises' product prices and emission reduction behaviors under different allocation rules (Zhang et. al., 2015). Research on the driving factors behind carbon dioxide emission changes in China can inform better carbon emission reduction policies and help develop a low-carbon economy. Using 2005–2010 data (Wang et. al., 2015) find that economic development was the largest factor of increasing carbon dioxide emissions. It is essential to incorporate the environmental objective in the transportation mode selection problem as transportation is a main contributor to carbon emissions (Chen et. al., 2016). These findings do not only help firms to make optimal decisions under different carbon emission reduction policies but also support policy makers to develop effective policies on carbon emissions reduction. (Ji et. al., 2017) analyze a detailed model which incorporates both cap-and-trade regulation and consumers' low-carbon preference. Two emission reduction strategies are compared, including single manufacturer's emission reduction in production strategy and joint emission reduction strategy, of which entails manufacturer's and retailer's emission reduction.   To give policy-making insights to governments as well as production and carbon emission reduction decision-making insights to manufacturers (Cao et. al., 2017) explore which policy is better for society. The results show that the carbon emission reduction level increases as the carbon trading price increases, whereas it is independent of the unit low carbon subsidy.   To better promote supply chain emission reduction (Sun et. al., 2020) analyze the carbon emission transfer and emission reduction problem among enterprises within the supply chain, integrating the influence of government emission reduction policies and the low carbon market. Notably, under decentralized decision-making, when the emission reduction technology lag time and consumers' low carbon preference remain within a specific range, the carbon emission transfer behavior exerts a positive promoting effect on the emission reduction of the supply chain.   Carbon emission efficiency is an important indicator that can be used to measure progress toward carbon emission reduction targets. The relationship between green technology innovation and carbon emission efficiency has not been adequately studied, and the transmission mechanism is not yet clear. Based on the above research gaps, taking 32 developed countries that have proposed carbon neutral targets as research samples (Dong et. al., 2022) adopt spatial mediation model and spatial moderation model to analyze the transmission effects of economic development, urbanization, and financial development on environment-related green technology and carbon emission efficiency.

In this paper, we explore innovative strategies aimed at optimizing supply chain networks for industrial carbon emission reduction. Our work addresses an urgent need to reduce the environmental footprint of current industrial operations and offers a pathway for a sustainable industrial environment. The work conducted includes:

1. **Environmental Impact Evaluation**: We began by establishing baseline measurements of carbon emissions resulting from various supply chain stages, ranging from raw material procurement to the production phase, from distribution to customer usage and end-of-life recycling or disposal.

2. **Modeling Framework Development**: The development of a novel framework incorporating strategic planning and optimization of supply chain networks was the next step. We created a multi-objective optimization model that balanced operational efficiency and environmental sustainability, including the reduction of carbon emissions.

3. **Adaptive Carbon Emissions Indexing (ACEI)**: We introduced an innovative methodology called Adaptive Carbon Emissions Indexing. This model analyzed real-time carbon emissions data, allowing for dynamic adaptations to supply chain strategies based on current emission levels.

Through the course of this work, our research assists industries in transitioning towards a sustainable future, where operational efficiency and environmental stewardiness are synergistically managed.

## 2. Carbon Emission Reduction

Industrial carbon emissions, dominantly resulting from the combustion of fossil fuels and subsequent release of carbon dioxide ($CO_2$), are pivotal contributors to the escalating issue of climate change. In this discussion, we undertake an analysis of carbon emissions, emphasizing their significance in an industrial context, to underscore the urgency for interventions aimed at emissions reductions.

Carbon emissions refer to the release of greenhouse gases (GHGs), primarily $CO_2$, into the environment as a result of human activities. Industries, as significant consumers of energy, are often associated with high carbon emissions stemming from their production processes, energy usage, transportation requirements, and waste disposal practices. Most industrial processes rely on the burning of fossil fuels for energy. When fossil fuels (coal, oil, and natural gas) burn, they release $CO_2$ along with other pollutants. Furthermore, carbon emissions also arise from activities like deforestation and land-use changes that the industrial sector often precipitates. These activities reduce the capacity of our ecosystems to absorb $CO_2$, thereby exacerbating the net increase of atmospheric carbon.

Industries contribute to approximately 21% of global GHG emissions, thus playing a cardinal role in the trajectory of climate change. Within industries, energy-intensive sectors such as power production, manufacturing, and transportation are the predominant sources. Unabated, these emissions will continue to propel temperature rise, jeopardizing ecosystem stability and human health, and causing irreversible damage to our planet. However, industry facilitates economic growth, employment, and societal development, and therefore its function is essential. Hence, the quandary lies in balancing the economic contributions of industry with the environmental crisis unfolding due to carbon emissions.

The industrial sector has unrivaled leverage in driving carbon emissions reduction. Given its enormous share in global GHG emissions, even small improvements achieved in energy and process efficiency can result in significant overall reductions. These improvements can be catalyzed through technological advancements and redesigns in process workflows, alongside shifts towards cleaner energy sources and conscious resource utilization. Moreover, regulatory bodies worldwide are enacting stringent emission regulations, and industries that fail to comply will face significant financial implications in the form of penalties and fines. On the positive side, organizations that successfully reduce their carbon footprints could gain considerable competitive advantage, reputation boost, and access to increasingly conscious consumers and investors. Moreover, the need for carbon emission reduction extends beyond the immediate criterion of environmental benefits. It influences a multitude of factors, including public health, socioeconomic structure, geopolitical stability, and most importantly, the sustained existence of life on Earth.

In the grand scheme, carbon emissions stand at the crossroads of industrial growth and environmental stability. Through intelligent decision-making and strategic investments, industries can significantly curtail their carbon emissions. Prompt action towards reduction and eventual neutrality in carbon emissions can potentially defer the staggering threats posed by climate change. Therefore, understanding the concept and implications of carbon emissions within the industry sector forms the groundwork to transition towards a more sustainable future. Beyond the realm of environmental responsibility and risk management, the mitigation of carbon emissions can lead industries to the stronghold of resilience, profitability, and global citizenry.

As we grapple with the dual mandate of socio-economic growth and planetary conservation, the attempts to curtail industrial carbon emissions position us at a critical juncture in this struggle to reclaim our Earth's future.

## 3. Supply Chain Network Optimization

The supply chain network encompasses the entire journey of a product, from raw materials sourcing to production, to distribution and end-consumer delivery, and finally to the disposal or recycling of used products. Supply chain network optimization entails employing methodologies, tools, and technologies to achieve the most efficient and effective network configuration.

An optimized supply chain network rationalizes operational activities, minimizes resource-intensive processes, streamlines workflows, reduces time-to-market, and lowers operational costs. Implemented correctly, it can boost the organization's overall performance, customer satisfaction, and competitive advantage.

Today's dynamic global economy and escalating environmental concerns necessitate a reorientation of traditional supply chain management. Industrial facilities must navigate their supply chain network management not only to maintain competitiveness but also to reduce their carbon footprint. The intersection of these two imperatives lies in supply chain network optimization aimed at carbon emission reduction. The typical industrial supply chain harbors many carbon-intensive activities, such as production processes, transportation, packaging, and waste management. Supply chain network optimization can significantly reduce these emissions by facilitating smart choices in sourcing materials, energy usage, logistical planning, and waste management. The improvement areas might include the adoption of cleaner energy sources in production, procurement of sustainable raw materials, consolidation of shipment loads, route optimization for logistics, efficient warehouse layouts to reduce energy consumption, and effective end-of-life product management strategies like recycling or upcycling. Significance of Supply Chain Network Optimization can be summarize as follows:

1. **Operational Efficiency and Cost Reduction**: An efficient supply chain network reduces redundant activities and waste. It decreases the usage of energy, material resources, and time, thus significantly lowering operational costs. Given the cost-intensive nature of many carbon reduction strategies, the cost savings achieved through optimization can offset the expenses and motivate industries toward environmental responsibility.

2. **Regulatory Compliance and Risk Avoidance**: As global regulators tighten environmental standards, industries face mounting pressure to align their operations with such standards. Supply chain network optimization can help meet and exceed these standards, avoiding potential fines and penalties associated with non-compliance.

3. **Reputation Enhancement**: As public awareness of climate change rises, more consumers and investors choose to associate with environmentally responsible organizations. Leveraging supply chain network optimization to reduce carbon emissions can significantly enhance a company's brand image, foster customer loyalty, and attract ethical investors.

4. **Supply Chain Resilience**: A well-optimized supply chain network is often more flexible and agile. It is better equipped to adapt to changes or recover from disruptions, aiding the overall resilience of the business. Given the increasing frequency and intensity of climate-related disruptions, this resilience could be vital for an organization's survival and growth.

5. **Innovation and Competitive Advantage**: Embracing supply chain network optimization for carbon emissions reduction often necessitates innovative solutions - such as clean technologies, renewable energy adoption, waste-to-wealth cycles, etc. These innovations can provide distinct competitive advantages in the market.

Supply Chain Network Optimization, targeted towards carbon emission reduction, challenges the notions of traditional industrial operations. It presses industries to rethink their supply chain models, innovate, and integrate sustainability at the core of their business strategy. This shift is not just a responsible move toward environmental stewardness, but also a business imperative in the backdrop of evolving regulatory landscapes, changing consumer preferences, and the urgent need to mitigate the impacts of climate change.

## 4. Adaptive Carbon Emissions Indexing-based Supply Chain Optimization

Addressing the burgeoning issue of industrial carbon emissions necessitates strategic approaches that can dexterously integrate organizational profitability and environmental sustainability. To this end, we propose an innovative strategy based on Adaptive Carbon Emissions Indexing (ACEI) for supply chain optimization. Herein, we elaborate on the conceptualization, development, and implications of this groundbreaking approach. Understanding the pressing need for industries to adapt and develop sustainable practices, our strategy begins with quantifying and categorizing carbon emissions within the supply chain. The Adaptive Carbon Emissions Indexing is a novel mechanism designed to perform this function. With this tool, carbon emissions at each node and process of the supply chain can be quantified, indexed, and tracked in real-time.

The inclusion of adaptive algorithms within this indexing mechanism affords dynamic responses to generated data, subsequently triggering adjustments to supply chain strategies. Essentially, this tool offers a granular and quantifiable measure of an organization's carbon footprint, woven into its supply chain operations, serving as a key performance indicator. With the ACEI serving as the data bedrock, the development of the optimization strategy pivots on building a carbon-efficient supply chain. The crucial steps encompass data interpretation, strategy formulation, execution, and iterative refinements, all underpinned by machine learning algorithms for robust and dynamic performance:

1. **Data Interpretation**: The ACEI offers a comprehensive overview of carbon emissions data from various processes across the supply chain. Techniques like data clustering and feature extraction, aided by machine learning, help map high-emission hotspots.

2. **Strategy Formulation**: Leveraging the ACEI data insights, strategies are designed to address elevated carbon emission zones. These strategies might include changes in sourcing materials, modifications to production processes, adjustments to logistical operations, or enhancements in waste management practices.

3. **Execution**: Deploying the formulated strategies necessitates synchronizing changes across the supply chain network. This step requires a clear alignment of goals, effective communication, and structured project management.

4. **Iterative Refinements**: Continually fed real-time data, the ACEI helps monitor the effectiveness of the executed strategies, thus enabling regular feedback, verification, and iterative improvements.

Moreover, the proposed ACEI-based optimization strategy is a powerful tool equipped to drive industrial carbon emissions reduction, which can be summarized as follows:

1. **Informative Decision-Making**: The ACEI provides quantifiable, real-time carbon emissions data that can guide informed, data-based decision-making, bypassing the hurdle of estimations or assumptions.

2. **Dynamic Adaptability**: The adaptive nature of this strategy ensures high responsiveness towards environmental regulations, market trends, and technological advancements, thus maintaining its relevance and efficiency over time.

3. **Operational Efficiency**: By highlighting carbon-intensive processes, the strategy enables targeted improvements, which often coincide with enhancing operational and cost efficiencies, providing an incentive for industries to adopt this strategic approach.

4. **Environmental & Regulatory Compliance**: This strategy serves as a robust tool for carbon management, helping industries align with stringent regulatory standards and contribute positively to global carbon reduction targets.

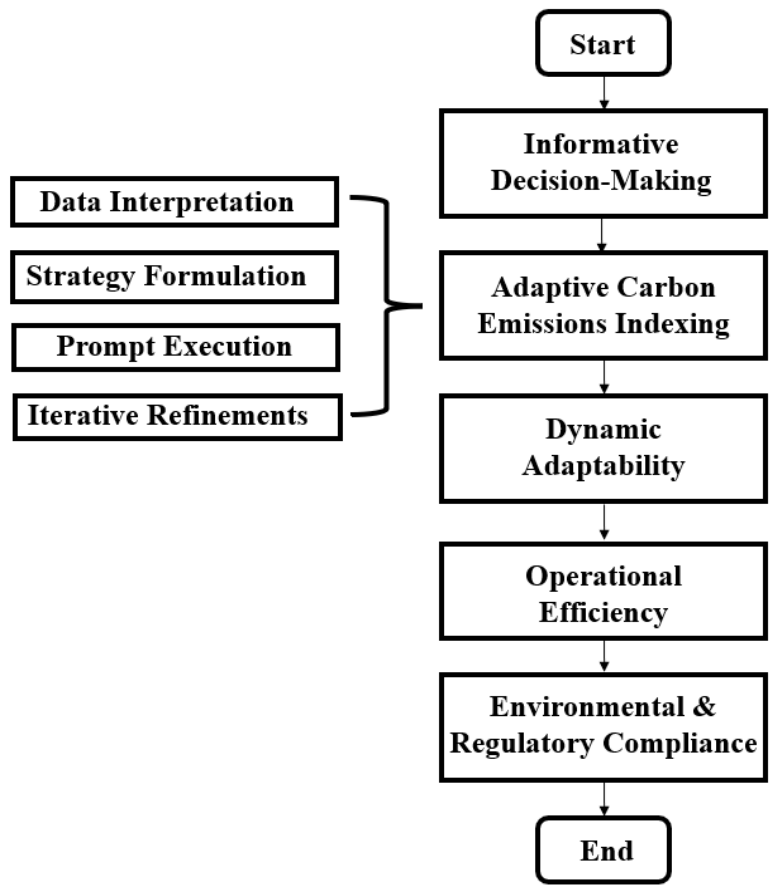

**Figure** 1: Flow chat of ACEI-based Supply Chain Optimization

Our proposed strategy based on Adaptive Carbon Emissions Indexing for supply chain optimization combines advanced technology with strategic environmental responsibility. The intelligent integration of machine learning algorithms empowers this tool to deliver robust, dynamic, and adaptive performance. In effect, it drives industrial supply chains to embrace sustainability, while securing economic gains, encapsulating the dual mandate of modern businesses. As a tool for deciphering and tackling the complex issue of carbon emissions, the ACEI-based optimization strategy offers a promising pathway toward a sustainable industrial future.

**5. Case study**

In this section, we investigate a manufacturing corporation called GreenTech Manufacturing Corporation to showcase the procedure of proposed method. GreenTech Manufacturing Corporation, a leading global player in manufacturing advanced technology solutions, presents a compelling case of striving towards carbon neutrality. In response to the mounting pressure to curb their carbon emissions, GreenTech embarked on a transformational journey of supply chain optimization, employing our proposed strategy based on Adaptive Carbon Emissions Indexing (ACEI). The richness of this narrative lies in the intricate challenges faced, and the trailblazing initiatives employed to overcome them.

*4.1 Identification of Problem:*

GreenTech, facing wavering profit margins and increasingly stringent carbon emissions regulation, recognized the need to reengineer its supply chain operations. A comprehensive audit revealed higher carbon emissions at key stages of its supply chain, including raw materials sourcing, production, logistics, and waste disposal. To address these challenges, GreenTech decided to incorporate the ACEI-based supply chain optimization strategy to reduce its carbon footprint while enhancing operational efficiency.

*4.2 Implementation of ACEI-based Optimization Strategy*

GreenTech started by integrating the ACEI into their digital infrastructure, enabling real-time tracking, quantifying, and indexing of carbon emissions across their expansive supply chain. It unraveled CO2-intensive hotspots, specifically in raw materials sourcing and production. The detailed data provided by ACEI, combined with machine learning capabilities, allowed for quantifiable decision-making and dynamic strategy development.

1. **Materials Sourcing & Production**: GreenTech switched to procuring raw materials from suppliers committed to environmentally friendly practices, reducing upstream carbon emissions. Concurrently, they modified production processes to incorporate energy-efficient equipment and practices further driving down emissions.

2. **Logistics**: ACEI revealed the transportation segment to be another significant emissions contributor. GreenTech optimized its logistics by consolidating shipments, optimizing routes, and transitioning part of their fleet to electric vehicles.

3. **Waste Management**: With a higher focus on their product's end-of-life impact, they initiated a take-back policy, encouraging consumers to return used products for appropriate disposal or recycling, thereby curbing disposal-associated emissions.

*4.3 Evaluation of the Implemented Strategy*

Monitoring outcomes via the ACEI data streams reflected a substantial reduction in carbon emissions within two years of implementing the strategy. The highest reduction was noted in the sourcing and production stages, indicating successful strategic implementation. Additionally, GreenTech documented an improvement in operational efficiency by 23% and a considerable reduction in operating costs.

Moreover, the brand reputation of GreenTech soared, with a noteworthy increase in customer loyalty and shareholder confidence. GreenTech's endeavors exemplified their commitment to sustainability, resulting in their favorable ranking in several global environmental performance indices.

*4.4 Case Scenario and Results*

GreenTech's case provides striking evidence of the efficacy of an ACEI-based supply chain optimization strategy. The company mitigated not only its environmental footprint but also enhanced its financial and operational performance, thus proving the viability of integrating sustainability into the quarters of supply chain management. GreenTech's story underscores the powerful potential of ACEI-based supply chain optimization strategies, demonstrating how companies can streamline their supply chain practices, reduce their carbon emissions, and enhance their market position. GreenTech's strides in reducing its carbon footprint reflect a maturing acknowledgment in the industrial sector that economic growth and responsible environmental stewardship are not mutually exclusive. Instead, a company's ability to leverage efficient strategies for carbon emission reduction can serve as a critical foundation for their competitiveness, resilience, and overall corporate sustainability in the face of a rapidly changing global business landscape.

## 6. Discussions

Certainly, we have made promising strides towards integrating efficient strategies on supply chain network optimization for industrial carbon emissions reduction. However, challenges undeniably remain. Key among these is the data conundrum. Comprehensive, accurate, and real-time data are of paramount importance for ACEI-based strategies. However, the capturing, processing, and standardizing of such data are often rife with complexities. Data availability varies across industries and regions, and an absence of globally standardized procedures for data collection further exacerbates this issue. Additionally, deploying an ACEI-based system entails technological sophistication and computational power, which may serve as potential stumbling blocks for industries with limited resources. The shifting nature of carbon emission regulations across different countries poses yet another roadblock, necessitating continual adaptations in supply chain networks. Resistance from within organizations, especially stemming from concerns over the significant initial capital investment, also needs to be negotiated.

To forge a pathway into promising future developments, three key focus areas emerge. The first revolves around enhancing technological capabilities and developing resource-efficient ACEI-based software that can be deployed across businesses of various sizes and economic capacities. Concurrently, efforts should pivot on creating rigorous and universally recognized methodologies for emissions data collection, processing, and standardization, ideally leveraging technological advancements to assist in accurate and consistent data integration. Lastly, the corporate culture of organization needs to evolve to foster acceptance and drive the deployment of such data-driven decarbonizing strategies. Clear evidence of financial returns on their investments, alongside demonstrable environmental benefits, would greatly aid in promoting this shift. Additionally, consistent global policies and regulations regarding carbon emissions should be pursued to ensure seamless strategy adaptation across international markets. Thus, while the path ahead presents hurdles, it also provides a fertile ground for research and development, vowing a brighter and more sustainable future for industrial supply chain management.

## 7. Conclusion

In conclusion, the push for Industrial Carbon Emission Reduction necessitates a paradigm shift in supply chain operations. Through our research, we have proposed an innovative strategy based on Adaptive Carbon Emissions Indexing for effective supply chain network optimization. Harnessing the power of data and advanced technology, this strategy enables organizations to capture, analyze, and respond dynamically to their carbon footprint data across supply chain networks. The potential of this approach is immense, not only for achieving tangible reduction in carbon emissions, but also for optimizing cost and operational efficiency, enhancing brand reputation, complying with environmental regulations, and fuelling sustainable innovation. In the face of the dual mandate of economic growth and environmental stewardship, our proposed strategy provides an actionable route aligned with future-forward industrial activities. As the industry grapples with increasing complexities amidst climate change and evolving markets, efficient strategies for supply chain optimization, underlined by carbon emission reduction, offer a critical pathway for the industry to stride towards sustainable and profitable operations. Our research underscores the integration of these strategies, developing the groundwork necessary for industries to transition towards a greener and more sustainable future.